\UseRawInputEncoding
\documentclass[aps,twocolumn,groupedaddress,superscriptaddress,floatfix,pra]{revtex4-1}
\usepackage{amsmath}
\usepackage{amssymb}
\usepackage{graphicx}
\usepackage{textcomp}
\usepackage{bm}	
\usepackage{soul}
\usepackage{graphicx,subfigure}

\usepackage[usenames,dvipsnames]{color}

\usepackage[braket,qm]{qcircuit}

\begin{document}

\title{Effective Hamiltonian for Stokes--anti-Stokes pair generation with pump and probe polarized modes}

\author{R. Acosta Diaz}\email{racosta@if.ufrj.br}
\affiliation{Instituto de F\'isica, Universidade Federal do Rio de Janeiro, CP68528, Rio de Janeiro, Rio 
de Janeiro 21941-972, Brazil}

\author{C.H. Monken}
\affiliation{Departamento de Física, ICEx, Universidade Federal de Minas Gerais. Av. Antonio Carlos, 6627, 
Belo Horizonte, MG, 31270-901, Brazil}

\author{A. Jorio}
\affiliation{Departamento de Física, ICEx, Universidade Federal de Minas Gerais. Av. Antonio Carlos, 6627, 
Belo Horizonte, MG, 31270-901, Brazil}

\author{Marcelo F. Santos}\email{mfsantos@if.ufrj.br}
\affiliation{Instituto de F\'isica, Universidade Federal do Rio de Janeiro, CP68528, Rio de Janeiro, Rio 
de Janeiro 21941-972, Brazil}

%\pacs{xxxx, xxxx, xxxx}                                         
\date{\today}

\begin{abstract}
In the correlated Stokes-anti--Stokes scattering (SAS) an incident photon interacts with a Raman-active 
material, creating a Stokes photon and exciting a quantum vibrational mode in the medium, which is 
posteriorly annihilated on contact with a second incident 
photon, producing in turn an anti-Stokes photon. This can be accomplished by real and virtual 
processes. In real process the quantum mode shared between the Stokes and anti-Stokes events is a real 
particle, whereas in virtual processes the pair formation is mediated by the exchange of virtual particles. 
Here, we introduce a Hamiltonian to describe the pair production in SAS scattering, for both types 
of process, when stimulated by two orthogonally polarized laser pulses in a pump-and-probe configuration.
We also model the effect of the natural decay of the vibration created in the Stokes event and compute the probability 
of producing SAS pairs. Additionally, we follow the dynamics of the vibration by considering the Stokes and anti-Stokes fields as external reservoirs, obtaining thus a master equation for the reduced density matrix for the vibrational population. Finally, we compare our theoretical results with recently published experimental data.
\end{abstract}

\maketitle

\section{INTRODUCTION}
The Raman spectrum of materials arises as a consequence of the inelastic scattering of light by matter  
\cite{Ra} which is underlaid on two types of process: the Stokes process S, characterized by the 
annihilation of an incoming photon of frequency $\omega_L$ and the creation of a redshifted one, of 
frequency $\omega_S$ together with the excitation of a quantum vibrational mode of frequency $\nu$ in the 
target medium; and the anti-Stokes process AS, where the incident photon and an existing vibration are 
annihilated, generating a blueshifted photon of frequency $\omega_{AS}$. In each event, energy is 
conserved such that the frequency of the outgoing photon is equal to the frequency of the pump plus (AS) 
or minus (S) that of the vibration. As proposed by Klyshko \cite{SaS,Kl}, an additional process is also 
possible: the correlated Raman scattering, or Stokes-anti-Stokes process (SAS). In this 
case, the overall scattering involves the annihilation of two incoming photons of frequency $\omega_L$ 
and the creation of a pair composed of one Stokes and one anti-Stokes photon, such that 
$\omega_S + \omega_{AS} = 2 \omega_L$. The non-classical nature of the correlated Stokes--anti-Stokes 
Raman scattering components has been demonstrated experimentally in several materials, including graphene~\cite{Jo0,Jo1}, diamond~\cite{Lee,MKa,And} and other transparent media, including water and other liquids~\cite{Saraiva,FSA1,MKa2}. Interestingly, this phenomenon has raised connections with other condensend matter fields, such as superconductivity~\cite{Saraiva} and phonon-pumping effect induced by strong fields in surface-enhanced Raman spectroscopy~\cite{kneipp2000surface,le2006vibrational}.

Such correlated SAS pairs can involve the exchange of a real vibration, created in a Stokes process and 
subsequently destroyed by an anti-Stokes one. In this case, not only does the above mentioned overall 
energy conservation holds, but the pair is created red and blue shifted with respect to 
the incoming laser energy by the vibrational mode frequency, 
i.e. each Stokes and anti-Stokes photon has a well defined frequency $\omega_{AS,S} = \omega_L \pm \nu$. 
These pairs, here addressed as \textit{real} SAS, lie at the core of recent pump-and-probe experiments 
exploiting the vibrations of the material as a potential quantum memory, and are characterised by a time 
correlation dominated by the lifetime $\tau_{_{P}}$ of such vibrational modes. In solids, these vibrations 
are phonons of typical lifetimes of a few picoseconds~ \cite{Lee,And,Vele,Aguiar} and have led to new 
experiments to store and process classical and quantum information on picosecond time-scales at room 
temperature with several different materials~\cite{Lee,van,Re1,Re2,Eng1,Bu,Eng2,Fis}.  

The correlated Raman scattering also takes place out of resonance, as demonstrated in Ref. 
\cite{Saraiva} in many different materials such as diamond and diverse liquids. 
In the denominated \textit{virtual} processes, the energy of S and AS events are tuned 
out of resonance with the vibrational mode, so that the generation of photon pairs happens by means of the 
exchange of virtual vibrations. The photon pair produced by a virtual process can be viewed as the 
photonic analogue of the Cooper pairs in superconductivity \cite{BCS}. Some properties of the so-called 
photonic Cooper pairs, or PCP, were studied in~\cite{Agu,FSA1} for open air propagation and their 
existence has also been predicted for waveguides in~\cite{YZ}. While the real SAS process has a 
characteristic timescale dictated by the phonon lifetime, as previously observed in different studies, in 
the virtual SAS process the exchange of virtual phonons is expected to be nearly instantaneous, 
therefore limited only by the inverse bandwidth of the excitation pump pulse. This scattering time difference was experimentally demonstrated in~\cite{Aguiar} in a time-delayed pump and probe setup with orthogonally polarized pulses. 

All these experiments use either a microscopic or a phenomenological model always designed specifically for the tested regime (real or virtual exchange of phonons). A full microscopic model able to fully describe not only both extreme regimes but also intermediate ones was still missing.

In this paper, we use an extension of the model proposed in~\cite{Parra} to provide a unified theoretical 
framework for the creation of pairs through either virtual or real vibration exchange. We calculate the 
time evolution of the overall quantum state of both photonic and vibrational fields and obtain the 
probability distribution of SAS pairs generation considering the natural decay of the vibrations in the 
material. We test our model analyzing the experimental data produced in~\cite{Aguiar} where both regimes are investigated in diamond.
In particular, the phenomenological model used to explain the resonance (real phonons) data in~\cite{Aguiar} is based on the decay of the
phonon population between pump and probe pulses. We also consider the Stokes and anti-Stokes fields as creating and absorbing reservoirs for the 
vibrations and derive a master equation for its dynamics. This allows us to test the validity of this phenomenological model.

\section{General theoretical framework}
Our stating point is the effective Hamiltonian proposed in Ref.~\cite{Parra}, given by
\begin{eqnarray} \label{H1}
 H&=&\hbar\omega_{L}a^{\dagger}a + \hbar\nu c^{\dagger}c + \hbar\omega_{S}b^{\dagger}_{S}b_{S} + 
   \hbar\omega_{AS}b^{\dagger}_{AS}b_{AS} \nonumber \\
   &+& \hbar\lambda_{S}\left(ab^{\dagger}_{S}c^{\dagger} + 
   H.c\right) + \hbar\lambda_{AS}\left(ab_{AS}^{\dagger}c + H.c\right),
\end{eqnarray}
where $b^{\dagger}_{S}$ $(b_{S})$, $b^{\dagger}_{AS}$ $(b_{AS})$, $c^{\dagger}$ $(c)$ and $a^{\dagger}$ 
$(a)$ stand for the creation (annihilation) operator of S, AS, phonon and incident fields, respectively. The constant 
$\lambda_{S}$ $(\lambda_{AS})$ denotes the coupling between the laser and the material, responsible for 
the Stokes (anti-Stokes) events. The Stokes and anti-Stokes photon frequencies are given by 
$\omega_{S,AS}=\omega_{L}\mp\nu$, for real processes, with $\omega_{L}$ and $\nu$ being the 
pump and phonon frequencies, respectively. The Hamiltonian is obtained by handling the Raman scattering as an 
optical parametric amplification process~\cite{FG}. It is valid within the coherence time of the pumping 
laser, whether continuous or pulsed. Our procedure is also similar to the one used in Ref.~\cite{Vele}.

\subsection{Full dynamics for the quantum fields}
In~(\ref{H1}), the pump laser is assumed to be a quantum field but, given its very large power and the relatively low
count of Raman photons, its depletion can be ignored and it can be replaced by a classical function of time. We are, then, left with 
three quantum fields whose dynamics we proceed to calculate, the Stokes and anti-Stokes photonic modes and the phonon of the material. 

First, we extend Hamiltonian~(\ref{H1}) to the case in which the SAS process is stimulated by the incidence of two orthogonally 
polarized laser pulses, and the Stokes and anti-Stokes fields can be generated at arbitrary frequencies. This choice of the polarization of the pulses is motivated by the experimental data used to test our model but does not represent any limitation to the model itself. The total Hamiltonian describing the dynamics of pair generation, then, reads
\begin{equation} \label{TotalHamiltonian}
 H(t)=\hbar\nu c^{\dagger}c+H_{0_{H}}+H_{0_{V}}+H_{I_{H}}(t)+H_{I_{V}}(t).
\end{equation}

The first term is, once again, the free energy of the vibrational mode, whereas the second and third terms, given by 
\begin{eqnarray}
 H_{0_{H(V)}}&=&\int_{0}^{\infty}d\omega \hbar\omega
  b^{\dagger}_{S_{H(V)}}(\omega)b_{S_{H(V)}}(\omega) \, \nonumber \\
  &+& \, \int_{0}^{\infty}d\omega
  \hbar\omega b^{\dagger}_{AS_{H(V)}}(\omega)b_{AS_{H(V)}}(\omega), 
\end{eqnarray}
are the free energy of the photonic Stokes and anti-Stokes fields for the two orthogonal polarizations of the pump. Finally, the two last terms describe the coupling of all the fields via the material, and read
\begin{eqnarray} \label{HI1}
 H_{I_{H}}(t)&=&\int_{0}^{\infty}d\omega \hbar g(\omega)f_{H}(t-t_{0})b_{S_{H}}^{\dagger}(\omega)c^{\dagger} \nonumber  \\
 &+&  \int_{0}^{\infty}d\omega \hbar g(\omega)f_{H}(t-t_{0})b_{AS_{H}}^{\dagger}(\omega)c  + h.c.
\end{eqnarray}

\begin{eqnarray} \label{HI2}
 H_{I_{V}}(t)&=&\int_{0}^{\infty}d\omega \hbar g(\omega)f_{V}(t-t_{1})b_{S_{V}}^{\dagger}(\omega)c^{\dagger}  \nonumber \\
&+&  \int_{0}^{\infty}d\omega \hbar g(\omega)f_{V}(t-t_{1})b_{AS_{V}}^{\dagger}(\omega)c  + h.c.
\end{eqnarray}
where $g(\omega)$ gives the coupling between pump, created photon at frequency $\omega$ and the vibrational mode. $f_{j}(t-t_j)$ describes the amplitude of the pump field and we assume that the converted photons preserve the polarization of the pump field, a hypothesis justified by experimental data (see Ref.~\cite{Aguiar} for example). 

So far, the model is generic and allows for any type of pulses sent into the material. However, because we are particularly interested in comparing our results with previous experiments, we are going to choose Gaussian profiles $f_j(t-t_i)=\alpha_j e^{-(\frac{t-t_i}{\sigma})^2}e^{-i\omega_Lt}$ for the time dependence of the pump ($i=0$, $j=H$) and the probe ($i=1$, $j=V$), where $j=H,V$ denotes the polarization of the pulse and $\delta \tau = t_1-t_0$ is the time delay between them. The coupling terms then become (in the interaction picture):
\begin{eqnarray}
 \bar{H}_{I_{H}}=\hbar \alpha_{H}e^{-\left(\frac{t-t_{0}}{\sigma}\right)^{2}}\left[
 \int_{0}^{\infty}d\omega g(\omega)e^{i\Delta_{1}t}b^{\dagger}_{S_{H}}(\omega)c^{\dagger}\right. & &
                                                                                \nonumber \\
 + \left.\int_{0}^{\infty}d\sigma g(\sigma)e^{i\Delta_{2}t}b^{\dagger}_{AS_{H}}(\sigma)c\right]  + h.c.
                                                                                                 & &,                                                                         
\label{HI}
\end{eqnarray}
\begin{eqnarray}
 \bar{H}_{I_{V}}=\hbar \alpha_{V}e^{-\left(\frac{t-t_{1}}{\sigma}\right)^{2}}\left[
 \int_{0}^{\infty}d\omega g(\omega)e^{i\Delta_{1}t}b^{\dagger}_{S_{V}}(\omega)c^{\dagger}\right. & & 
                                                                                    \nonumber \\ 
 + \left.\int_{0}^{\infty}d\sigma g(\sigma)e^{i\Delta_{2}t}b^{\dagger}_{AS_{V}}(\sigma)c\right]+ h.c.
                                                                                                 & &,  
\label{HV}                                                                                                
\end{eqnarray}
where $\Delta_{1}=\omega_{L}-\omega-\nu$ and $\Delta_{2}=\omega_{L}-\sigma+\nu$.

After being created, the photonic fields propagate freely and will be eventually collected by the detectors. The phonon, however, can decay, e.g. due to its interaction with other vibrations of the material. Therefore, its dynamics, and by consistency, that of the entire system, is not unitary and cannot be properly described by solving the Schroedinger equation with Hamiltonian~(\ref{TotalHamiltonian}). In order to take the vibrational decay into consideration, we calculate the dynamics of the system by solving the following master equation (in the interaction picture):
\begin{equation} \label{MasterEq}
 \frac{d\rho}{dt}=-\frac{i}{\hbar}\left[\bar{H}_{I_{H}}(t)+\bar{H}_{I_{V}}(t),\rho\right]
 +\mathcal{L}(\rho), 
\end{equation}
where $\rho$ is the density operator for all the quantum fields (Stokes, anti-Stokes and phonon) and the Lindblad term $\mathcal{L}(\rho)$ is given by
\begin{equation} \label{LimTer}
 \mathcal{L}(\rho)=\gamma\left(2c\rho c^{\dagger}-c^{\dagger}c\rho
                           -\rho c^{\dagger}c\right),
\end{equation}
with $\gamma$ being the decay rate (proportional to the inverse of the lifetime, $\gamma = \tau_{_{P}}^{-1}$) of the vibrational field. This equation assumes a dissipative channel at zero temperature which is a good approximation for most experiments where the vibration corresponds to an optical phonon, whose average number of thermal excitations at room temperature is of the order of $10^{-3}$. The integration in time of Eq.~(\ref{MasterEq}) gives the general solution for our system. A similar approach for the vibrational dynamics in the context of Raman scattering was used in~\cite{ACSNano}. There, however, the scattering involves plasmons in a cavity whereas here it happens in free space.

\subsection{Master equation for the phonon population} \label{MasEqSec}
Eq.~(\ref{MasterEq}) contains all the information about the dynamics of each quantum field involved in the
scattering process, as well as any eventual time correlation among them. Therefore, its integration in time should provide a proper description of the production of SAS pairs for any frequency displacement $\Delta$. However, as suggested by the phenomenological approach used in~\cite{Aguiar}, the time-delayed cross-correlation function of SAS pairs can be obtained essentially from the time dependence of the vibrational population. The underlying assumptions are that the real SAS processes that mostly contribute for the pair counting are those were the correlation is driven solely by the shared phonon (i.e. the individual scattering processes are statistically independent) and that each event is very rare. In other words, the Stokes process is spontaneous (the vibrational field is basically in the vacuum at the arrival of each pump pulse), in each pulse at most one phonon is excited (Stokes processes are rare), and when the anti-Stokes process takes place it is solely due to the phonon created within the same pump-probe pair. All these assumptions meet the experimental conditions. First, the time distance between pairs of pump-probe pulses (13 ns) is much larger than the decay time of phonons in the material (few picoseconds) and, as mentioned before, the average number of thermal phonons is very low (around $10^{-3}$), thus justifying the first assumption of pump pulses reaching the material in its vacuum state of phonons. Second, the rate of Stokes photons is of the order of $10^4$ counts per second, while the pulses strike the material at around 76 MHz, which means that each pulse has a probability smaller than $10^{-3}$ of creating a Stokes photon and, hence, a phonon. Finally, in order to guarantee the absolute statistical independence of the real SAS pairs, the detection post-selects Stokes and anti-Stokes photons of orthogonal polarizations, each sharing the polarization respectively of the pump and the probe pulses. That guarantees that the Stokes comes from the pump and the anti-Stokes from the probe.

Fortunately, the robustness of the model described in the previous subsection also relies on the fact that it allows to derive an equation exclusively for the dynamics of the phonons themselves. As we proceed to show, under the circumstances of performed pump and probe experiments, when discussing the phonons alone, Stokes and anti-Stokes fields play the role of external reservoirs. The Stokes field will correspond to a pumping reservoir that incoherently creates phonons in the material, while the anti-Stokes field will enhance the dissipation rate of such phonons. Given the typical time scales of the pulses and the weakness of individual scattering processes, we will proceed to derive a master equation for the phonons taking into account these extra reservoirs. As it will become clear soon, the time dependence of the pulse will reflect in time dependent dissipative or pumping rates.

To get an expression for the master equation in the Lindblad form, according to the scenario just 
described, we take as a ground the second-order contribution to the evolution of the reduced phonon 
density operator $\rho_{_{Ph}}=\mathrm{Tr}_{_{SAS}}\rho$ in the time convolutionless 
approximation~\cite{Liv},

{\small{
\begin{equation} \label{CTL}
 \frac{d\rho_{_{Ph}}}{d t}=-\lambda^{2}\int_{t_{i}}^{t}du\,\mathrm{Tr}_{_{SAS}}
 \left\lbrace\left[\bar{H}_{I}(t),\left[\bar{H}_{I}(u),\rho_{_{Ph}}(t)\otimes
  \rho_{_{SAS}}\right]\right]\right\rbrace,
\end{equation} 
}}

\noindent
where $\rho_{_{SAS}}$ is the density operator of the SAS field. The constant $\lambda$ stands for 
the strength of the interaction that in our case is represented by the couplings $\alpha_{_{H(V)}}g$. In 
order to derive the final equation, we will assume that these couplings follow two conditions:
\begin{equation} \label{Cond}
 |\alpha_{_{H(V)}}g(\omega)|\ll\omega \quad \mathrm{and}
                    \quad\left|g(\omega=\omega_{L}\mp\nu)\right|^{2}\sim g_{0}^{2}.
\end{equation}
Signs $(-)$ and $(+)$ correspond to S and AS events, respectively. The first condition implies the weak coupling regime which is fully justified in our procedure. $|\alpha g|^2 \sim 10^{4} \textrm{Hz}$ is proportional to the count of Stokes photons and, therefore, much smaller than the optical frequency of the fields at $\omega \sim 5\times 10^{14}\textrm{Hz}$, which is of the order of the correlation time of our photonic reservoirs. The second condition means that the response of the material ($|g(\omega)|$) is basically flat in the range of frequencies involved in the experiment, which is valid when considering materials with electronic gap much higher then the excitation energies involved, such as in diamond. With no loss of generality, we will approximate it by a Gaussian function 
\begin{equation} \label{gw}
 |g(\omega)|^{2}=g_{0}^{2}\textsl{e}^{-\frac{(\omega-\omega_{L})^{2}}{\delta\omega^{2}}},
\end{equation} 
such that $\delta\omega$ is taken much greater than all the physical parameters involved in the process. Finally, each pulse is still well resolved in frequency, ${1}/{\sigma}\ll\omega_L$. When put together, these conditions allow us to derive a master equation governing the phonon population distribution as result of the interaction with the SAS field. Assuming the experimentally verified equal intensity for the pump and probe pulses, $|\alpha_{{H}}|^{2}=|\alpha_{{V}}|^{2}=\alpha_{0}^{2}$, the contribution of the Stokes and anti-Stokes reservoirs for the dynamics of the phonons reads (see the Appendix for details):
\begin{eqnarray} \label{EqMestre3}
\frac{d\rho_{_{\mathrm{Ph}}}}{d t}&=&2\pi \left(\alpha_{0}g_{0}\right)^{2}
        [e^{-2\left(\frac{t-t_{0}}{\sigma}\right)^{2}}+ e^{-2\left(\frac{t-t_{1}}{\sigma}\right)^{2}}] \nonumber \\
        & &\times[2c^{\dagger}\rho_{_{\mathrm{Ph}}}c 
        +2c\rho_{_{\mathrm{Ph}}}c^{\dagger} - \{\{c,c^\dagger\},\rho_{_{\mathrm{Ph}}}\}],
\end{eqnarray}
where $\{A,B\}=AB+BA$. 
The overall phononic evolution is obtained by adding the dissipative Lindblad term, in the form 
of Eq. (\ref{LimTer}), to the right hand side of Eq. (\ref{EqMestre3}). This extra term accounts for the dissipation due to other phonons of the material, as previously mentioned.

\section{RESULTS}
\subsection{Overall behaviour as a function of $\Delta$}
In principle, the model described by Eq. (8) allows us to obtain the time correlation $P_{SAS}(\delta \tau)$ of SAS photons at any pair of detection frequencies, i.e. for any $\Delta$, and as a function of any delay time $\delta \tau$ between the pump and probe pulses. In order to calculate $P_{SAS}(\delta t)$ we numerically integrate Eq. (8) for a given set $\{\Delta,\delta \tau\}$. This integration can be computationally demanding if the required Hilbert space is too large. However, in the regime of operation of most SAS experiments, including all the ones mentioned in the introduction, the scattering probabilities, both of a single Stokes photon and a vibration ($\sim 10^{-4}$) or a SAS pair ($\sim 10^{-7}$), are very low, and so is the mean number of thermal phonons. That means that for each pair of pump and probe pulses, the most probable event, by far, is that all the quantum fields (Stokes, anti-Stokes and phonon) start in the vacuum and remain at the same state throughout the evolution. Sometimes a Stokes photon and a phonon are produced and more rarely a SAS pair is produced. This allows us to truncate the Hilbert space to a small basis, thus significantly reducing the computational time. The size of this basis depends on the particular values of the coupling constants, system frequencies and decay rates. In general, we have chosen a large enough basis to guarantee that the probability of the higher excited states remained at least three orders of magnitude lower than the probability of forming a SAS pair.

To be more specific, firstly, we obtain the density matrix $\rho$ solving Eq. (8) numerically by considering a truncated basis of the 
Hilbert space whose elements represent the possible states that occur in an unlikely single SAS event. They 
can be represented in a generic way by a vector $\ket{S,AS,P_{h}}$, where S, AS and $P_h$ stand for 
the number of particles in the Stokes, anti-Stokes and phonon fields, respectively. The elements composing 
the basis set includes the vectors with $S\,(AS \, \textsl{and} \, P_{h})=0,1,2$. To choose the truncated basis, we use an 
extended set in the most favourable configuration (largest probability of pair formation) and confirm that larger occupation numbers 
have very low probability of showing up. Before the incidence of the first pulse, all three fields are assumed to be empty, i.e. the initial state for the 
integration is $\rho(t_i)=\ket{0,0,0}\bra{0,0,0}$ and the time delay $\delta t$ is fixed in a range between $-2$ and $12$ ps. 
Subsequently, the probability to create a SAS photon pair for that particular $\delta t$, $P_{SAS}(\delta t)$ is obtained from $\rho$ 
by calculating $P_{SAS}(\delta t)=\mathrm{Tr}\left\lbrace\ket{1,1,0}\bra{1,1,0}\rho\right\rbrace$. It is pertinent to note that the lower and upper limits of integration are taken respectively at $t_i=-3$ ps and $t_f=14$ ps. That guarantees that
both pump and probe pulses, Gaussian shaped of width smaller than $1$ ps, are properly encompassed in the dynamics for all of the values of the time 
delay between them as implemented in the experiment whose results we describe in this work. Also note that, due to computational limitations (the involved numbers may become too small), extending the limits of integration may give rise to errors in the numerical procedure.

\begin{figure}[ht!]
\includegraphics[scale=0.36]{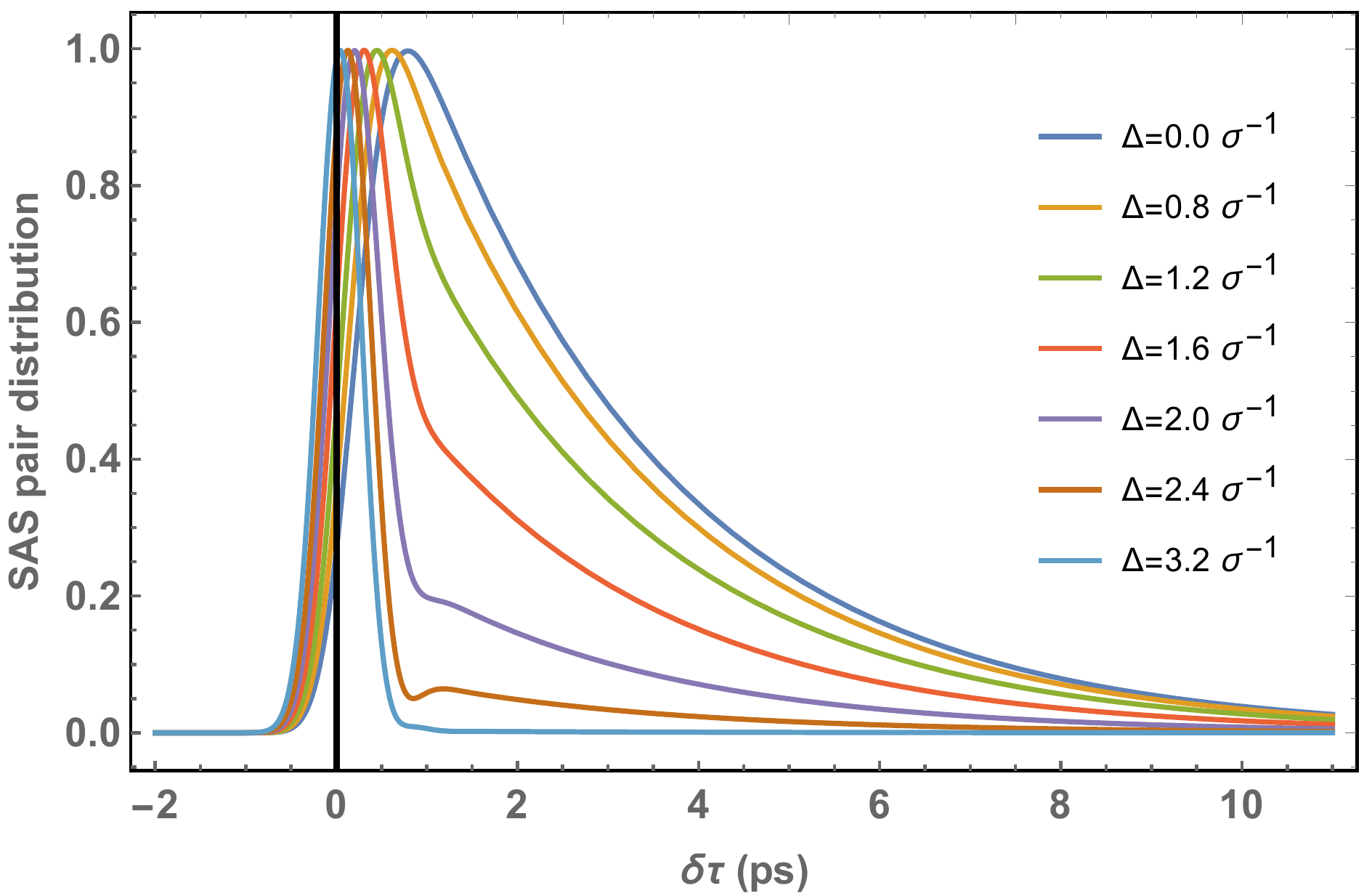}
\caption{Normalised SAS pair distribution $P_{SAS}(\delta \tau)$ as a function of the delay $\delta \tau$ between the pump and probe pulses, for different values of the detuning $\Delta$. Real processes are dominant for $\Delta=0$ while the virtual ones are preponderant for large values of $\Delta$. Each pulse is a Gaussian function of time with FWHM $\sigma =0.40$ ps. $\Delta$ is given in unities of $\sigma^{-1}$. The vibration dissipation rate is $\gamma = 0.36 \times 10^{12}\textrm{s}^{-1}$ corresponding to a phonon decay time of $\tau_{_{P}} = 2.78$ ps.}  
\label{all_regimes_a}
\end{figure}

We present in Fig.~\ref{all_regimes_a}, the normalised distribution $P_{SAS}(\delta \tau)/P_{Max}$ as a function of the delay between the pulses for different values of $\Delta$. One clearly sees the expected transition of regimes. For $\Delta = 0$, the time correlation function shows two characteristic features observed in different experiments made at the Raman resonance peaks: a smooth decay influenced by the dissipation time of the vibration and a shift from zero delay of its maximum value. This confirms that this regime is dominated by real SAS processes. The shift from zero is physically justified by the fact that for real processes, it is more efficient to wait for the pump pulse to generate the Stokes photon and the phonon before the probe reaches the sample to produce the anti-Stokes photon. Naturally, this shift cannot be too large because the dissipation of the vibration to other channels (scattering against other phonons of the material, for example) is a competing deleterious effect. At the other extreme, for very large $\Delta$, the width of the curve is determined basically by the width of the pulses and the most efficient pair production takes place at zero delay ($\delta \tau =0$). This indicates that in this regime the dominating process is the virtual SAS, in which case the best scenario happens when both pump and probe coincide in the material, increasing the chance of a simultaneous creation of the Stokes and anti-Stokes photons. For intermediate frequency displacements, the curve is a mixture of the two processes with weights that depend on the specific value of $\Delta$, as expected.

\begin{figure}[ht!]
\subfigure[]{
\includegraphics[scale=0.41]{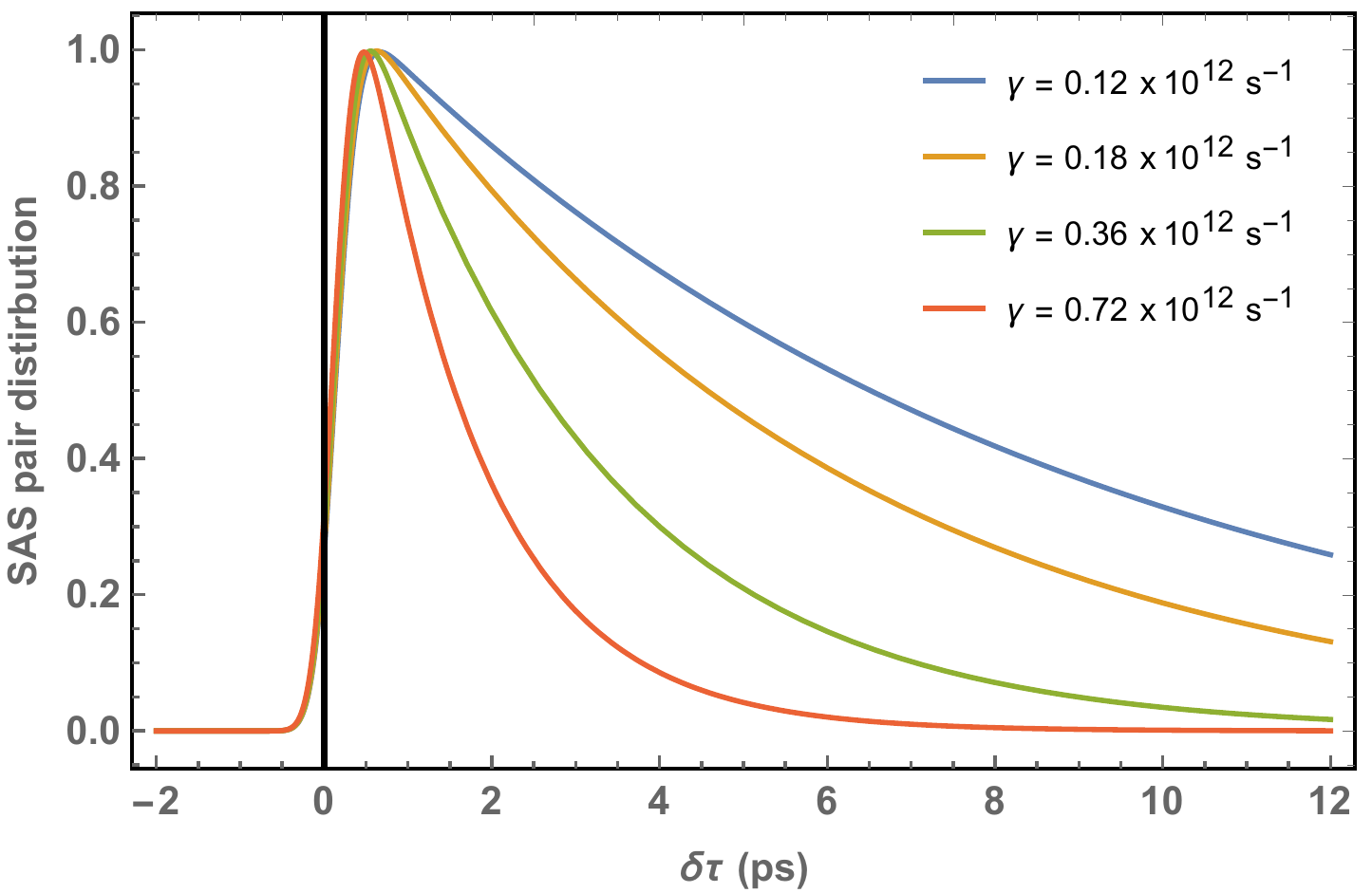}\label{all_regimes_b}
}
\subfigure[]{
  \includegraphics[scale=0.32]{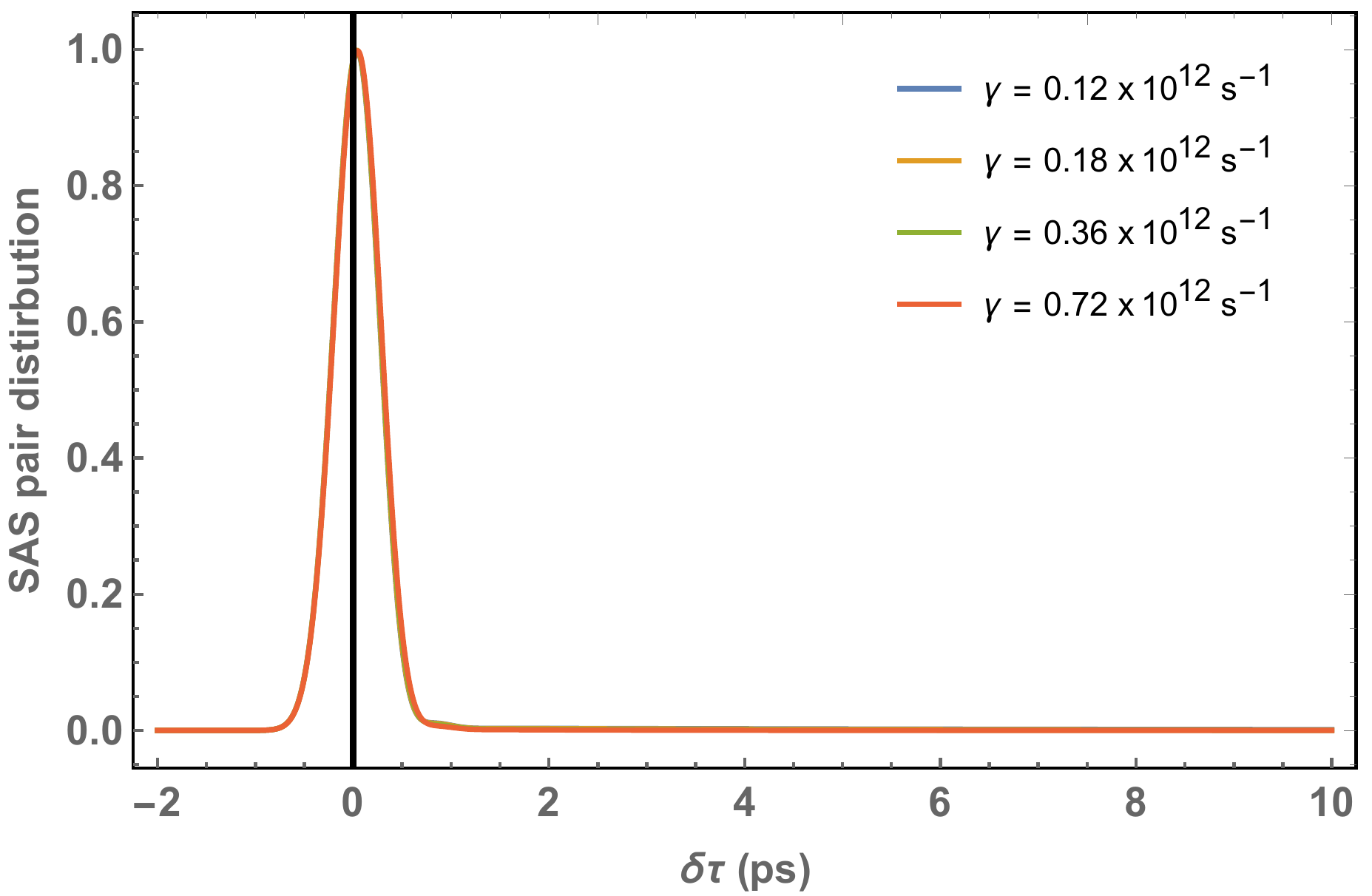}\label{all_regimes_c}
}  
\caption{Normalised SAS distribution $P_{SAS}(\delta \tau)$ for different values of the decay rate of the phonon population $\gamma$.  The values of $\gamma$ are given in unities of $10^{12} \textrm{s}^{-1}$ and the pulses are the same Gaussian functions of time used in Fig.1. (a) Real process ($\Delta = 0$). (b) Virtual process ($\Delta = 8$ THz). In this case, the curves for each value of $\gamma$ are superposed.}
\label{gammas}  
\end{figure}

In Fig.~\ref{all_regimes_b}, we show how the resonant curve ($\Delta =0$) depends on the dissipation time of the vibration. As expected, the faster the dissipation rate (larger values of $\gamma$) the tighter the time window for the formation of the pair. Also note that the maximum of the curve shifts slightly towards zero delay time for larger values of $\gamma$. This happens because the deleterious effect of the vibrational dissipation forces towards a slightly increased overlap between the two pulses in order to maximize the pair production. On the other hand, as displayed in Fig.~\ref{all_regimes_c}, in the large $\Delta$ scenario, the dissipation time of the vibration does not play any significant role, confirming that only virtual vibrations are exchanged. 

\subsection{Data analysis}
In order to test our model, we compare its predictions to the experimental results displayed in Ref.~\cite{Aguiar}. There, the authors performed a pump and probe experiment in diamond in the exact same framework of the theory developed here. They used two delayed cross-polarized pulse beams and a set of polarization selective detectors to count correlated Stokes and anti-Stokes photons pairs in two regimes, on and off the Raman resonance peaks. Previous theoretical and experimental results indicated that the former is dominated by the exchange of real phonons in the material, whereas the latter by the exchange of virtual vibrations, and the experiment was set to test these two configurations.

The first part of the experiment was used to establish that both S and AS photons carry majorly the same polarization of the excitation pulse. With this in mind, the authors proceeded with the time delayed cross-correlated pump-probe photon detections of SAS pairs. In each run, a $76$ MHz sequence of two pulses of orthogonal polarization, $H$ and $V$, were sent into the sample, each pulse centered at  $\omega_L = 632.8$\,nm and of Gaussian shape with FWHM $\sigma =0.40$ ps. A polarization dependent delay line placed before the sample established a switchable time separation $\delta \tau$ between the $H$ pump pulse and the $V$ probe pulse ranging from $-2$ to $+13$ ps. A dichroic mirror and polarizers were placed before the photon detectors to guarantee the collection of Stokes photons of $H$ polarization and anti-Stokes photons of $V$ polarization.

The detections were made at two different pairs of frequencies, $\omega_A$ and $\omega_B$. First, photons were detected at the Raman resonance peaks with detectors placed at $\omega_A = \omega_S = \omega_L-\nu$ and $\omega_B = \omega_{AS} = \omega_L+\nu$. At these frequencies, the pair generation process is dominated by real SAS. Then, the detection setup was adjusted to collect pairs at displaced frequencies $\omega_A = \omega_S +\Delta$, $\omega_B = \omega_{AS}-\Delta$, where $\Delta < \nu$ and the process is dominated by virtual SAS \cite{FSA1}. The displacement was made sufficiently larger than the width of the resonance peaks in order to guarantee the dominance of virtual processes.

In each case, they measured the SAS scattering intensity, i.e. the number of pair coincidence photon counts or, 
equivalently, the pairs of correlated photons detected simultaneously, by varying the time delay $\delta\tau$ 
between the $H$ pump pulse and the $V$ probe pulse. It was observed that the production rate of real SAS pairs decreases with the decay of the phonon population generated by the Stokes process. The measured lifetime of the phonon population was around 
$2.8$ ps, in agreement with results of other experiments. In contrast, in the virtual process, SAS pair production occurred primarily 
when the two laser pulses overlapped, indicating that it happened faster than the duration of a single pulse.

\begin{figure}[ht!]
\subfigure[]{
\includegraphics[scale=0.42]{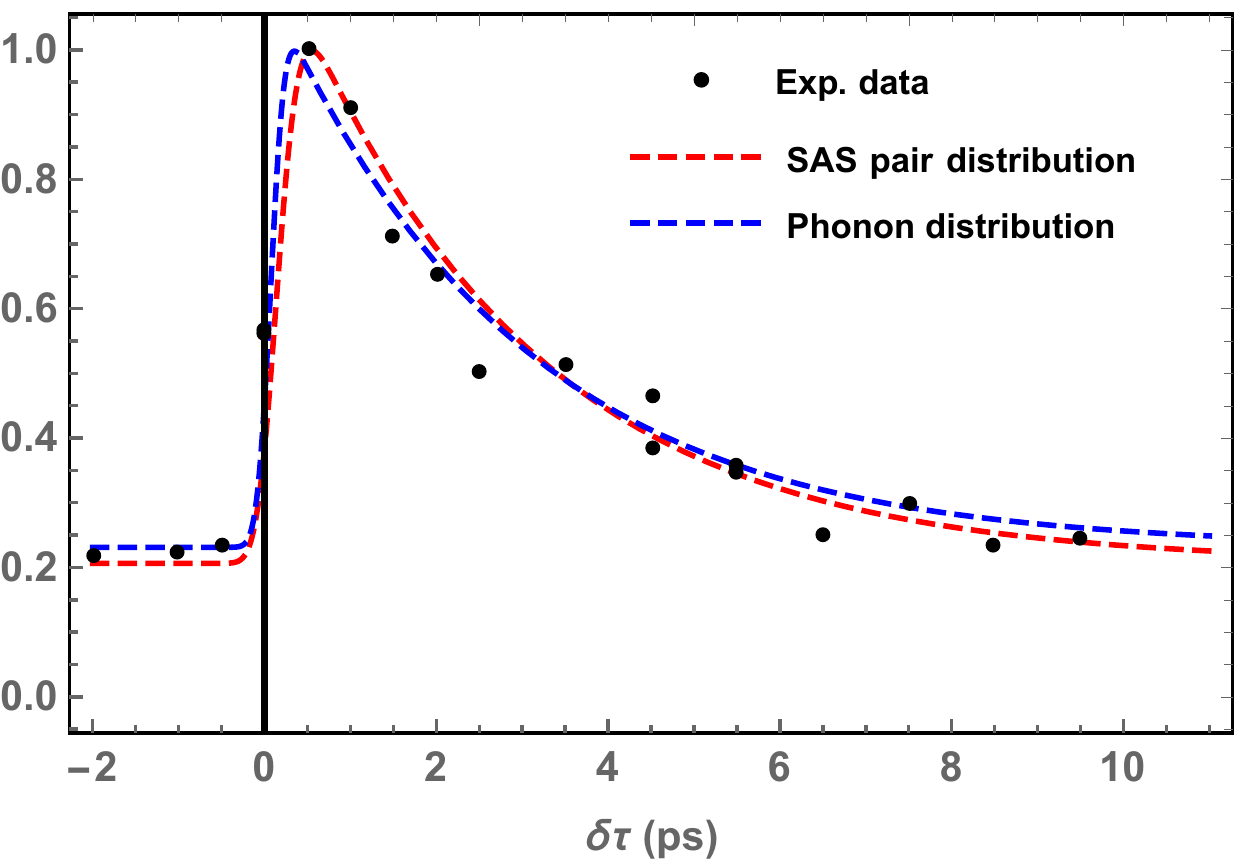}\label{RealSaS}
}

\subfigure[]{
  \includegraphics[scale=0.33]{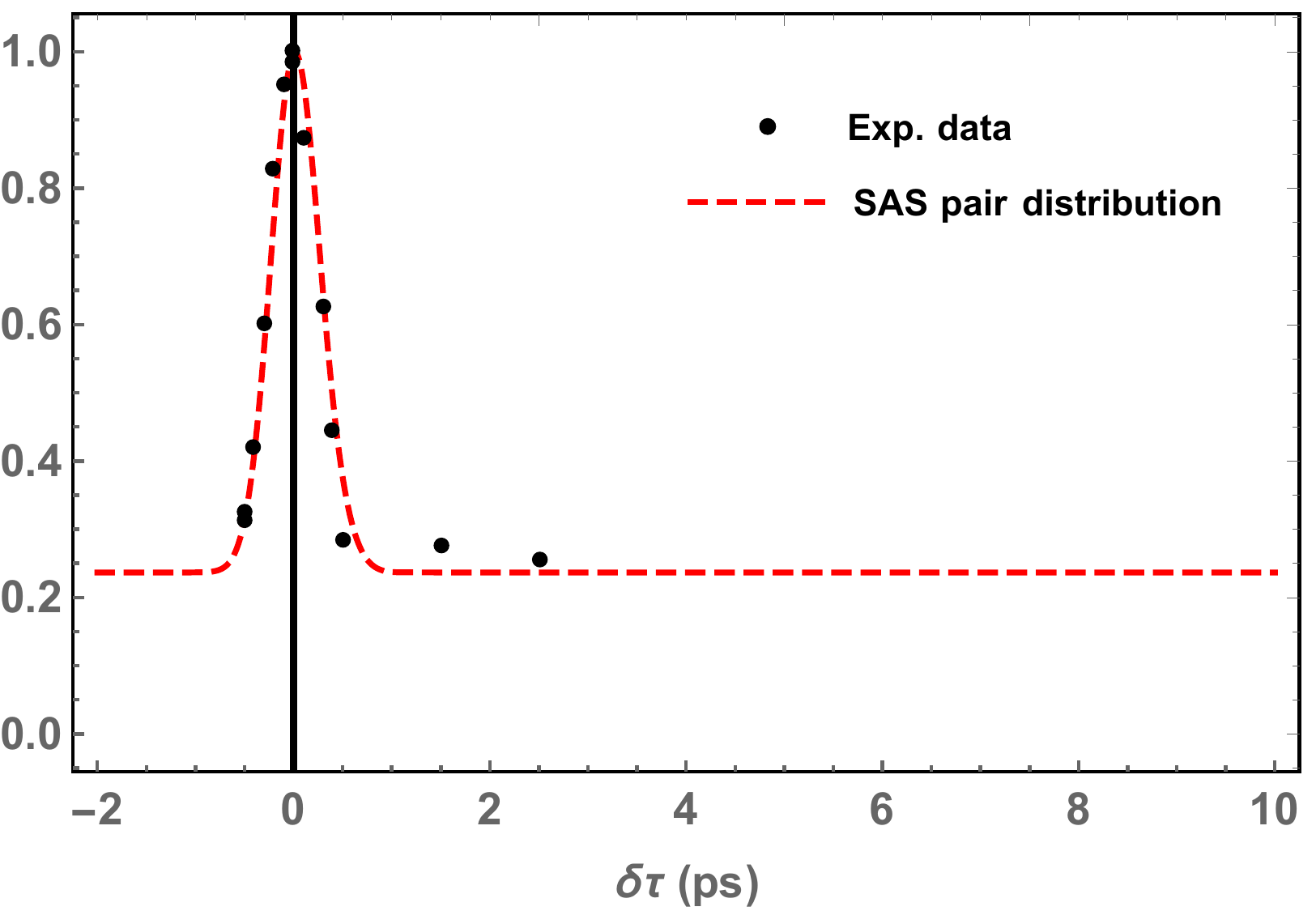}\label{VirSaS}
}  
\caption{The black dots in both plots correspond to experimental data published in~Ref. \cite{Aguiar}. (a) Real SAS. Red dashed line represents the probability of detecting a SAS pair from model IIA (full dynamics). Dashed-blue line corresponds to the probability of finding one vibration in the sample following model IIB (master equation for the vibrations). The values of the parameters involved are $\sigma=0.25$ ps, $\tau_{_{P}}=2.78$ ps e $C=0.25$. (b): Virtual process. The red-dashed curve corresponds to the probability of detecting a SAS pair from model IIA (full dynamics). The values of the parameters are $\sigma=0.4$ ps, 
$\tau_{_{P}}=2.78$ ps and $C=0.30$.}
\label{Fig}  
\end{figure}

In Fig.~\ref{Fig} we plot the experimental data from Ref.~\cite{Aguiar} and the theoretical curves obtained by solving Eq.~(\ref{MasterEq}) and calculating the normalised probability $P_{N_{SAS}}(\delta \tau)$ of finding a Stokes photon of polarization $H$ and an anti-Stokes photon of polarization $V$, at the respective detectors frequencies. The corresponding results for \textit{real} and \textit{virtual} processes are represented by the dashed--red lines in Fiqs.~\ref{RealSaS} and~\ref{VirSaS}, respectively.
The distribution is plot as a function of the delay $\delta \tau$ between pump and probe pulses, for values within the interval between $-2.0$ ps and $12.0$ ps, and it is given by 
\begin{equation}
\label{PSaS} 
P_{N_{_{SAS}}}(\delta\tau)=\frac{P_{_{SAS}}(\delta\tau)+CP_{_{max}}}{\left(1+C\right)P_{_{max}}}.
\end{equation}                                                                    
Note that this normalized function includes a shift to the minimum value of detected pairs, given by the constant $C$. This renormalization is necessary here because of experimental conditions not taken into account by the theoretical model. The model assumes perfect polarization in each pulse, also perfect polarization detection, no dark counts, no accidental counts and other experimental imperfections. All these effects contribute to a minimum pair counting present even for very large delays between the pulse and the probe. However, as it becomes clear, this simple correction, extracted from the experiment, is enough to properly adjust the experimental data. Also note that all the parameters used to draw the theoretical curves, both for the pulses and the lifetime of the phonon, are drawn from~\cite{Aguiar}.

Also note that, as mentioned in the previous section, the theory correctly predicts that at the Raman peaks (resonance), real pairs dominate and the probability of finding a pair depends on the decay of the phonon, whereas out of resonance the pairs are created only when the two pulses coincide, which is consistent with the formation of photonic Cooper pairs (PCPs), also observed in~\cite{Saraiva}, via the exchange of virtual phonons. 

In Fig.~\ref{RealSaS} we also plotted the theoretical curve for the phonon population, dashed-blue line, 
obtained by solving eq.~(\ref{EqMestre3}). Once again, the numerical integration was performed using a truncated basis in the Hilbert space of 
number states of phonons, $\ket{P_h}$, where $P_h=0,1,2$, justified, as before, by the fact that the chance of creating more than 2 phonons in each pair of pulses is very low and checked numerically by increasing the Hilbert space and confirming a very low probability of state $\ket{P_h=3}$ throughout the dynamics. As before, the system is initially in the vacuum state, $\rho_{P_h}(t_i)=\ket{0}\bra{0}$ and the region of integration is the same as in the
integration of Eq. (8). The probability of producing a phonon is given by 
$\rho_{P_h}=\mathrm{Tr}\left\lbrace\ket{1}\bra{1}\rho_{P_h}\right\rbrace$.

As in the previous case, the curve represents a normalized 
quantity:
\begin{equation} \label{Phdi}
 P_{N_{_{SAS}}}(\delta\tau)=\frac{P_{_{ph}}(\delta\tau)}{P_{ph_{max}}},
\end{equation}   
where $P_{_{ph}}(\delta\tau)$ is the probability to create a unique phonon in the 
%all 
entire process, while $P_{ph_{max}}$ is the maximum value for $P_{_{ph}}(\delta\tau)$.

The parameters used for the pulses, as well as the phonon natural lifetime are the same as in the dashed-red curve. Note that the result confirms the accuracy of the phenomenological model used in~\cite{Aguiar} for the probability of generating real SAS pairs. It also reinforces the validity of the hypothesis used to derive eq.~(\ref{EqMestre3}), that is: at the Raman peaks, the pair formation is indeed dominated by real processes and its dynamics depends basically on the probability of a second pump photon combining with the phonon created in the Stokes scattering in order to create the correlated anti-Stokes component of the pair.

\section{Conclusions}
In this paper we have introduced an effective Hamiltonian to describe the correlated SAS photon pair 
production in Raman scattering by both real and virtual processes. In particular, we considered the cases when the
material is shined by two laser pulses of orthogonal polarizations with a time-delay one from the other. 
We have also derived a non-unitary dynamics for the vibration of the material considering the Stokes and anti-Stokes
fields as external pumping and dissipative reservoirs. We tested the validity of our model by comparing the theoretical
results with the experimental data measured in Ref.~\cite{Aguiar}. The Hamiltonian model confirms the experimental data in both regimes (real and virtual pairs). In particular, our model predicts correctly that, at the Raman resonance peaks, the process is dominated by the lifetime of the phonon, therefore by real SAS pairs, whereas out of the resonance peaks, it is centered at zero delay and defined basically by the inverse linewidth of the pump and probe pulses, hence, dominated by virtual SAS pairs. Furthermore, the model for the dynamics of the phonon distribution also adequately describes the real SAS data. This second result demonstrates that the phenomenological model used in~\cite{Aguiar} is sound and correctly captures the essence of the production of pairs at the Raman resonance peaks and under the experimental conditions used therein.

The model here proposed should be general enough to be applied at a range of frequencies and for diverse materials that have not been experimentally tested yet. New experiments in this range and with different materials are desirable to either confirm or improve the model. Another interesting improvement that can be pursued is the generalization of the model for arbitrary polarizations of the pump and probe pulses as well as different spatial-temporal modes, including the addition of degrees of freedom that take into account the transversal profile of the pulses. This would overcome a limitation of the current model, allowing the description of more complex experiments that include measuring properties such as the orbital angular momentum of the generated pairs as well as their angular distribution.

\section{Acknowledgments}
We thank Filomeno S. de Aguiar J\'unior for providing us with the experimental data. This work was 
supported by FAPERJ Projects No. E-26/202.290/2018 and No. E-26/202.576/2019 and by CNPq Projects No. 
302872/2019-1, 307481/2013-1, 429165/2018-8 and No. INCT-IQ 465469/2014-0.

\section{Appendix} \label{App}
Here we derive the master equation in the Lindblad form of the reduced phonon density operator, eq.~(\ref{EqMestre3}). To this aim, we consider the second order contribution of the dynamics of an open system in the time convolutionless approximation, which is defined by 
eq.~(\ref{CTL}). First, let us remember that, since the photonic fields are initially in the vacuum, 
\begin{subequations}
\begin{equation} \label{cor1}
\textsl{Tr}_{SAS}\left\lbrace b_{S_{H(V)}}(\omega_{1})
b_{S_{H(V)}}^{\dagger}(\omega_{2})\rho_{SAS}\right\rbrace
=\delta(\omega_{1}-\omega_{2}),
\end{equation}
\begin{equation} \label{cor2}
\textsl{Tr}_{SAS}\left\lbrace b_{AS_{H(V)}}(\omega_{1})
b_{AS_{H(V)}}^{\dagger}(\omega_{2})\rho_{SAS}\right\rbrace
=\delta(\omega_{1}-\omega_{2}),
\end{equation}
\end{subequations}
and the remaining similar relations between the fields operators are all equal to zero.

Then, if we consider the Hamiltonians given by eqs.~(\ref{HI})-(\ref{HV}), we get for Eq.~\ref{CTL} the following expression:
\begin{eqnarray} \label{meq}
  \frac{d\rho_{_{Ph}}}{d t}= & - & \int_{t_{i}}^{t}du\bigg[f_{H}(t-t_{0})f_{H}(u-t_{0})  \nonumber \\ 
                             & + & f_{V}(t-t_{1})f_{V}(u-t_{1})  \bigg]\times  \nonumber \\
  & & \left[\, \int_{0}^{\infty}d\omega|g(\omega)|^{2}e^{i\Delta_{1}(u-t)}cc^{\dagger}\rho_{_{Ph}}
                                                               \right. \nonumber \\
  & + & \int_{0}^{\infty}d\omega|g(\omega)|^{2}e^{i\Delta_{2}(u-t)}c^{\dagger}c\rho_{_{Ph}}
                                                                        \nonumber \\
  & + & \int_{0}^{\infty}d\omega|g(\omega)|^{2}e^{-i\Delta_{1}(u-t)}\rho_{_{Ph}}cc^{\dagger}
                                                                        \nonumber \\
  & + & \int_{0}^{\infty}d\omega|g(\omega)|^{2}e^{-i\Delta_{2}(u-t)}\rho_{_{Ph}}c^{\dagger}c 
                                                                       \nonumber \\ 
  & - & \int_{0}^{\infty}d\omega|g(\omega)|^{2}e^{-i\Delta_{1}(u-t)}c\rho_{_{Ph}}c^{\dagger}
                                                                         \nonumber \\
  & - & \int_{0}^{\infty}d\omega|g(\omega)|^{2}e^{-i\Delta_{2}(u-t)}c^{\dagger}\rho_{_{Ph}}c
                                                                         \nonumber \\
  & - & \int_{0}^{\infty}d\omega|g(\omega)|^{2}e^{i\Delta_{1}(u-t)}c\rho_{_{Ph}}c^{\dagger}
                                                                         \nonumber \\
  & - & \int_{0}^{\infty}d\omega|g(\omega)|^{2}e^{i\Delta_{2}(u-t)}c^{\dagger}\rho_{_{Ph}}c 
                                                                      \, \bigg],
\end{eqnarray}
The function $f_{j}(t-t_{j})$ in the integrand is given by
\begin{equation}
 f_{j}(t-t_{j})=\alpha_{j}e^{-\left(\frac{t-t_{j}}{\sigma}\right)^{2}},
\end{equation} 
such that, after removing the parenthesis in the integrand of (\ref{meq}), we have integrals of the 
following type,
\begin{equation}
 I=\int_{t_{i}}^{t}duf_{j}(u-t_{j})\int_{0}^{\infty}d\mu|g(\mu)|^{2}e^{\pm i\Delta_{b}(u-t)},
\end{equation} 
where the index $j$ replaces the $H$ and $V$ ones, and $t_{j}$ stands for the time where the maximum of 
the respective Gaussian occurs, while $\Delta_{b}=\Delta_{1,2}$. Now, since the integrand in the time 
integral decreases fairly quickly for times above and below of $t_{j}$ we extend the lower and upper 
integration limits to $-\infty$ and $+\infty$, respectively. This fact together with the assumptions 
established by Eqs. (\ref{cor1}-\ref{cor2}) determine the validity of the Markov approximation in our 
approach. Then, we can write,
\begin{eqnarray} \label{I}
 \!\!\!\!\!\!\!\!\!\!\!\!\!\!\!\!
 I & = & \alpha_{a}\int_{0}^{\infty}d\mu|g(\mu)|^{2}\int_{-\infty}^{\infty}du
   e^{-\left(\frac{u-t_{a}}{\sigma}\right)^{2}}e^{\pm\Delta_{b}(u-t)} \nonumber \\
   &  = & \alpha_{a}\sqrt{\pi}\sigma\int_{0}^{\infty}d\mu|g(\mu)|^{2}
         e^{-\left(\frac{\sigma}{2}\right)^{2}\Delta_{b}^{2}}e^{\mp i\Delta_{b}(t-t_{a})}.
\end{eqnarray} 
To perform the integral in frequency we assume that the response of the material $|g(\mu)|$ can be 
approximated by eq.~(\ref{gw}) as discussed before. In view of $\omega_{L}>>0$ we can again extend 
the lower integration, in (\ref{I}), to minus infinite, so the integral $I$ reads
\begin{widetext}
\begin{eqnarray} \label{I2}
 I & = & \alpha_{a}g_{0}^{2}\sqrt{\pi}\sigma\left[\int_{-\infty}^{\infty}d\mu 
   e^{-\frac{\left(\mu-\omega_{L}\right)^{2}}{\delta\mu^{2}}}
   e^{-\left(\frac{\sigma}{2}\right)^{2}\Delta_{b}^{2}}e^{\mp i\Delta_{b}(t-t_{a})}\right] \nonumber \\ 
   & = & \alpha_{a}g_{0}^{2}\sqrt{\pi}\sigma \left[e^{-\left(\frac{\omega_{L}}{\delta\mu}\right)^{2}}
         e^{-\left(\frac{s\sigma}{2}\right)^{2}}e^{\mp is(t-t_{a})}
         \int_{-\infty}^{\infty}d\mu e^{-A^{2}\mu^{2}}e^{B\mu}e^{\mp i(t-t_{a})\mu}\right] \nonumber \\
   & = & \alpha_{a}g_{0}^{2}\sqrt{\pi}\sigma \left[e^{-\left(\frac{\omega_{L}}{\delta\mu}\right)^{2}}
         e^{-\left(\frac{s\sigma}{2}\right)^{2}}e^{\mp is(t-t_{a})}\right]
         \left[\frac{\sqrt{\pi}}{A}e^{\frac{\big[B\mp i(t-t_{a})\big]^{2}}{4A^{2}}} \right],      
\end{eqnarray}
\end{widetext}
with
\begin{equation}
 \begin{cases}
   A^{2}=\left(\frac{\sigma}{2}\right)^{2}+\frac{1}{\delta\mu^{2}}, \\
   B=\frac{2\omega_{L}}{\delta\mu^{2}}-\frac{s\sigma^{2}}{2}, \\
   s=\nu-\omega_{L} \,\, (s=-\nu-\omega_{L}) \,\, \rightarrow \,\, S \,\, (AS).
 \end{cases}   
\end{equation}
Taking $\delta\mu^{2}$ much greater than all the physical parameters concerning to (\ref{I2}), it yields
\begin{equation}
 I=2\pi\alpha_{a}g_{0}^{2}e^{-\left(\frac{t-t_{a}}{\sigma}\right)^{2}}.
\end{equation}

\noindent
Finally, Eq.~(\ref{meq}) reduces to eq.~(\ref{EqMestre3}).

\end{document}